\documentclass{aip-cp}

\usepackage[numbers]{natbib}
\usepackage{rotating}
\usepackage{graphicx}

\begin{document}

\title{Low Space-Charge Intensity Beams in UMER via Collimation and Solenoid Focusing}

\author[aff1]{S. Bernal}
\author[aff1]{B. Beaudoin}
\author[aff1]{L. Dovlatyan}
\author[aff2]{S. Ehrenstein}
\author[aff1]{I. Haber}
\author[aff1]{R. A. Kishek}
\author[aff1]{E. Montgomery}
\author[aff1]{D. Sutter}
\affil[aff1]{Institute for Research in Electronics and Applied
Physics, University of Maryland, College Park, MD 20742}
\affil[aff2]{Case Western Reserve University, Cleveland, OH 44106}
\corresp[cor1]{Corresponding author: sabern@umd.edu}

\maketitle

\begin{abstract}
The University of Maryland Electron Ring (UMER) has operated
traditionally in the regime of strong space-charge dominated beam
transport, but small-current beams are desirable to significantly
reduce the direct (incoherent) space-charge tune shift as well as
the tune depression. This regime is of interest to model
space-charge effects in large proton and ion rings similar to
those used in nuclear physics and spallation neutron sources, and
also for nonlinear dynamics studies of lattices inspired on the
Integrable Optics Test Accelerator (IOTA). We review the
definition of space-charge intensity, show a comparison of
space-
charge parameters in UMER and other machines, and discuss a
simple method involving double collimation and solenoid focusing
for varying the space-charge intensity of the beam injected into
UMER.
\end{abstract}

\section{INTRODUCTION}
The initial motivation of the University of Maryland Electron Ring
(UMER) \cite{Kishek14} has been expanded recently to include
nonlinear dynamics studies in lattices inspired on the Integrable
Optics Test Accelerator (IOTA) at Fermilab \cite{Antipov17}, but
space-charge (SC) effects must be reduced significantly to conduct
``zero" beam-current control experiments \cite{Ruisard16}.
Furthermore, SC effects are increasingly important for new and
upgraded large storage and accumulator proton and ion rings, but
some of these are user facilities with limited time for beam
physics experiments. Examples of these machines are the Spallation
Neutron Source (SNS) at Oak Ridge national Laboratory, and the
heavy-ion synchrotrons (``SchwerIonenSynchrotron") SIS-18, and
SIS-100 in Germany. The tune shift from small (relative to the
 horizontal bare tune $\nu_{0x}$) direct (incoherent) SC effects is
given by \cite{Schindl06}

\begin{equation}
\Delta {\nu _x} = {\nu _{0x}} - {\nu _x} = \frac{{{r_{e,p}}}}{\pi
}\frac{N}{{{\beta ^2}{\gamma ^3}}}{\left[ {{\varepsilon _x}\left(
{1 + \sqrt {\frac{{{\varepsilon _y}{\nu _{0x}}}}{{{\varepsilon
_x}{\nu _{0y}}}}} } \right)} \right]^{ - 1}}\left(
{\frac{{{q_s}^2}}{A}} \right)\left( {\frac{{{F_x}{G_x}}}{{{B_f}}}}
\right),\
\end{equation}

\noindent where the first factor contains the classical radius of
the electron or proton, the second one the number of particles per
bunch $N$ and the relativity-theory parameters $\beta$ and
$\gamma$, and the third one the un-normalized, rms emittance
terms. The last two factors include the charge state $q_s$, the
mass number $A$, and parameters related to image forces ($F_x$),
type of distribution ($G_x$), and bunching factor ($B_f$). Over a
broad range of energies and beam currents in machines in
operation, the tune shift is $\Delta\nu_x$ = 0.1 –- 0.5. In UMER,
typically $\Delta\nu_x$ = 1.0 –- 5.0.

Further, ``beam intensity" is loosely defined as proportional to
$N$, with ``space-charge limited" meaning $\Delta\nu$ less than
0.5 (Laslett criterion \cite{Laslett63}). However, ``SC
intensity," $\chi$, as introduced independently by Reiser and
Davidson, is related to $\Delta\nu/\nu_{o}$, not just to $N$ or
$\Delta \nu$ \cite{Reiser08}. For small SC effects
\cite{Bernal16},

\begin{equation}
\chi  \simeq \frac{{2\Delta \nu }}{{{\nu _0}}} =
\frac{{K{R_m}}}{{\varepsilon {\nu _0}}},\,\,K =
\frac{I}{{{I_{0e,p}}}}\frac{2}{{{\beta ^3}{\gamma ^3}}}.
\end{equation}

In equation (2), $K$ is the beam perveance or SC parameter
\cite{Reiser08}, $R_m$ the machine average radius, and $I_{0e,p}$
is a characteristic current equal, approximately, to 17 kA for
electrons, and 31 MA for protons. For ions the characteristic
current is 31MA$\times$A, approximately, where $A$ is the mass
number. For the sake of simplicity, we have omitted the subscript
``$x$" in the tune in equation (2). Note also that the ratio
$R_m/\nu_0$ is the single-particle average betatron function.
Although increasing the operating tune would be an obvious way to
reduce SC intensity in UMER, the reduction would not be
substantial, because the tune would be limited to a maximum of
around 8.9 (standard $\nu_0$ is around 6.6.) Moreover, practical
problems would arise with overheating of the quadrupole magnets.

Another effect that arises from image forces and SC leads to a
{\it coherent} tune shift. This effect is ordinarily small at low
energies compared to the incoherent tune shift because it is
proportional to the ratio $(a_0/h)^2$, where $a_0$, $h$ are the
zero-current average radius, and the half-gap of the vacuum pipe,
respectively \cite{Bernal16}.

The {\it tune depression}, which is more commonly used than
$\chi$, is given by $\nu/\nu_0=(1-\chi)^{1/2}$, for arbitrary but
linear SC. Because of UMER's low energy (10 keV), $K$ is large
even for the smallest beam currents, which compensates for the
small machine radius to yield significant SC intensities. {\it
Both} the tune shift and the SC intensity $\chi$ are important for
characterizing beam dynamics: the former is naturally relevant to
betatron resonance crossing, while the latter can be related to
transverse beam stability in general \cite{Reiser08}. Despite
obvious differences in injection methods, beam loss mechanisms,
and bunch structure, UMER and the larger, planned or upgraded,
high-energy ion rings share key features: significant SC effects,
the absence of synchrotron radiation, and a modest number of turns
(100-1000). However, we must reduce the ratio $K/\epsilon$
significantly in UMER to approach the tune shifts and SC
intensities of the larger machines.

\section{SPACE CHARGE IN UMER AND OTHER MACHINES}
Table 1 summarizes parameters and SC calculations for 4 cases of
UMER operation, 3 cases of large, high-energy rings, and the IOTA
proton ring at Fermilab. The standard operation at 10 keV, 6 mA
(nominal) in UMER ($\nu_{0x}$ = 6.6) yields SC numbers close to
the corresponding ones for a possible heavy-ion fusion (HIF)
driver \cite{Barnard93}. Operation at 10 keV, 40 $\mu$A using the
DC electron gun method described in \cite{Bernal16b} yields a beam
with negligible SC parameters, as in a light source. Operation at
10 keV, 50 $\mu$A (see last section), would correspond roughly to
the SC regime of the proton accumulator ring at the Spallation
Neutron Source (SNS) at Oak Ridge National Laboratory at the end
of the accumulation cycle (1,000 turns). At the lowest end of SC
in UMER standard operation, we obtain SC parameters as in the IOTA
proton ring. The operation of UMER with lower beam currents
produced by double collimation and solenoid focusing is described
later.

\begin{table}[h]
\caption{Parameters relevant to direct space-charge in electron
and ion rings} \tabcolsep7pt\begin{tabular}{lcccc}
\hline
\tch{1}{c}{b}{Machine,\\ Circumference} & \tch{1}{c}{b}{Kinetic Energy,\\ $\beta=\upsilon /c$}  & \tch{1}{c}{b}{Peak Current, RMS\\ Emittance (morm.)}  & \tch{1}{c}{b}{$\nu/\nu_0$,\\ $\Delta\nu$}\\
\hline
UMER, 11.52 m & 10 keV, 0.195 & 6.0 mA, 1.3 $\mu$m$^*$   & 0.63, 2.4  \\
HIF $Xe^{+8}$ Driver, 429 m & 10 GeV, 0.381 & 1.0 kA, 50  m & 0.66, 2.1 \\
UMER, 11.52 m & 10 keV, 0.195 & 40 $\mu$A, 5.0 $\mu$m$^\dag$ & 1.00, 0.005  \\
ALS (LBNL), 197 m & 1.2 GeV, 1.000 & 400 mA, 3.5 nm & 1.00, 0.00 \\
UMER, 11.52 m & 10 keV, 0.195 & 50 $\mu$A, 0.07 $\mu$m$^\ddag$& 0.93, 0.45   \\
SNS Acc. Ring, 248 m & 1.0 GeV, 0.875 & 52 A, 120  m & 0.98, 0.15\\
UMER, 11.52 m & 10 keV, 0.195 & 0.6 mA, 0.4 $\mu$m & 0.86, 0.94   \\
IOTA Proton Ring, 40 m & 2.5 MeV, 0.073 & 8 mA, 0.3 $\mu$m & 0.96,
0.5 (1.2)$^\S$ \\
\hline
\end{tabular}
\end{table}
\vspace{-.25in}
\begin{quote}\begin{quote} {\small
*Typical operation.\newline \dag DC electron gun operation.
See \cite{Bernal16b}. \newline \ddag Double collimation (see Table
3).
\newline \S Unbunched (bunched) See \cite{Antipov17}.
 }
\end{quote}\end{quote}

Another important aspect of beam dynamics that justifies operation
with low SC-intensity beams in UMER is beam debunching. Without
longitudinal containment, coasting beams in UMER, initially
filling half the ring circumference, elongate until the two bunch
ends meet after a number of turns. A one-dimensional fluid model
allows us to calculate the ``sound" speed $C_S$, i.e., the speed
of charge rarefaction, and approximate debunching times according
to \cite{Bernal16, Sutter13}

\begin{equation}
\\{C_S} = \sqrt {\frac{{eg{\Lambda _0}}}{{4\pi {\varepsilon
_0}{\gamma ^5}{m_e}}}} ,\,\,\tau  = \frac{1}{{4{C_S}}}\left( {C -
{L_b}} \right).\
\end{equation}

The constant $\Lambda_0 = I_b/\beta c$ is the peak longitudinal
charge density, and $g$ is a form factor of order unity. $C$ is
the machine circumference (11.52 m) and $L_b$ the bunch's length
corresponding to 100 ns. Since the leading and trailing tails of
the bunch are not well defined, the debunching time as defined
above is only a reference parameter. Actually, the space charge
waves inside the bunch collapse at a time 2$\tau$ \cite{Sutter13};
further, beam current is still detected by the resistive
wall-current monitor in UMER for times of the order of 4$\tau$.
Table 2 summarizes results of longitudinal expansion. With
properly applied longitudinal focusing we can extend the number of
turns by a factor of 40 for 0.6 mA and 10 for 6 mA.

\begin{table}[h]
\caption{Beam debunching in UMER is greatly reduced at low
current.} \tabcolsep7pt\begin{tabular}{lcccc}
\hline \tch{1}{c}{b}{Beam Current\\} & \tch{1}{c}{b}{``Sound Speed"\\$C_S$ (m/s)} & \tch{1}{c}{b}{Approx. No. Turns\\ To Debunch}  & \tch{1}{c}{b}{Approx. No. Turns\\ With Long. Focusing}\\
\hline
60 $\mu$ A   & $10^5$  & 70   &  \\
0.6 mA  & 3$\times 10^5$ & 25 & 1,000   \\
6.0 mA       & 8$\times 10^5$ & 9 & 100   \\
21 mA        & $10^6$ & 6 &  \\
104 mA       & 2$\times 10^6$ & 3 &  \\
\hline
\end{tabular}
\end{table}

Independent control of current and transverse emittance is
desirable, but can only be partially realized in practice. We have
produced different beam currents with apertures located near the
electron gun output. Naturally, the beam current scales with the
area of the aperture, while the emittance scales with the aperture
radius. However, the smallest SC intensity achieved in this way
still leads to an incoherent SC tune shift of almost unity under
typical operating tunes in UMER. To achieve lower incoherent SC
tune shifts and intensities, ultra-low currents can be obtained
with three methods: DC electron-gun operation, photoemission, and
double collimation assisted with solenoid focusing. All three
methods are discussed briefly in \cite{Bernal16b}. In the next
section we describe in detail measurements and SC calculations
related to the third method, which we regard as the simplest one
to implement in UMER.

\begin{figure}
\begin{minipage}[c]{0.5\linewidth}
\includegraphics[width=\linewidth]{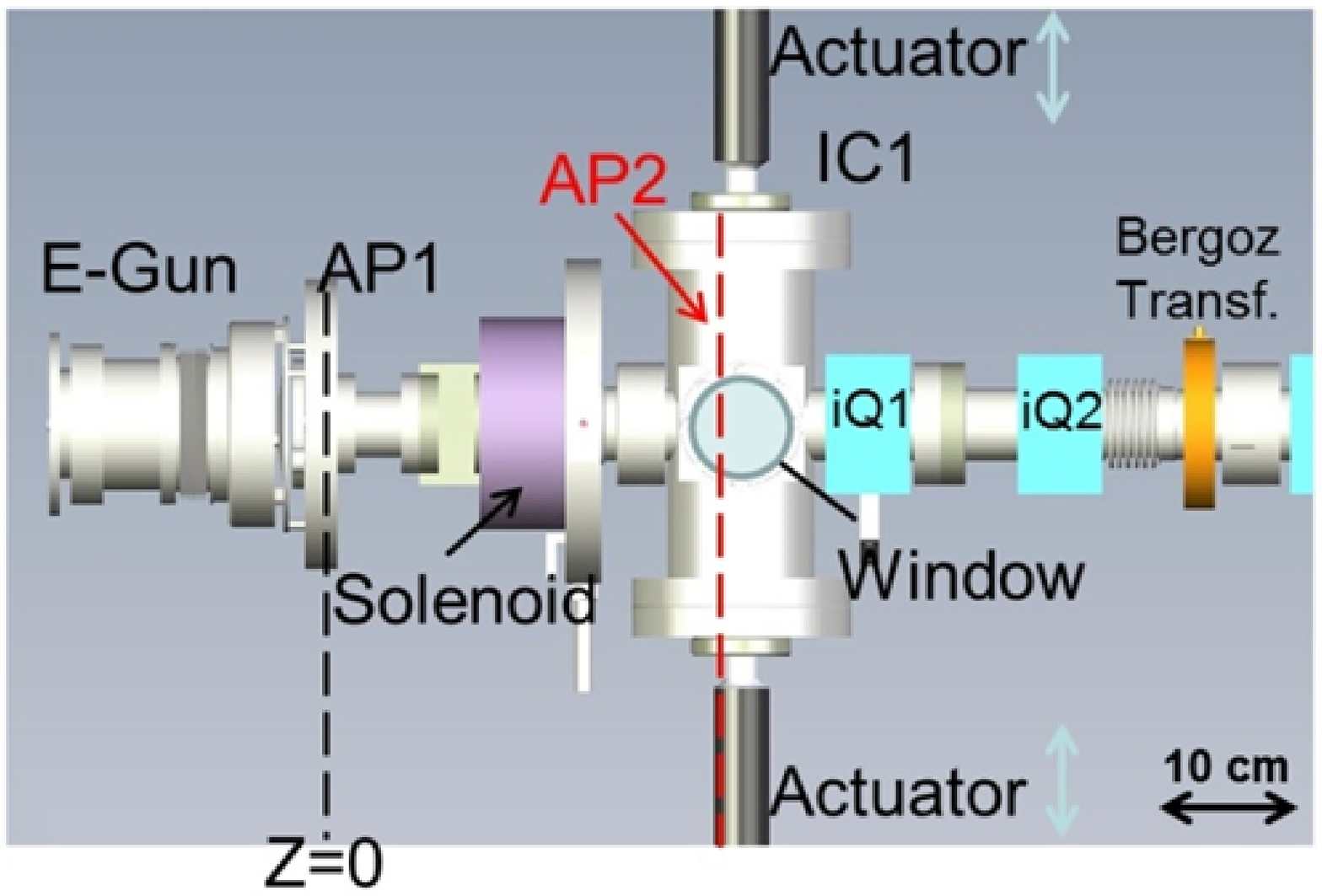}
\caption{}
\end{minipage}
\hfill \hspace{0.5in}
\begin{minipage}[c]{0.25\linewidth}
\includegraphics[width=\linewidth]{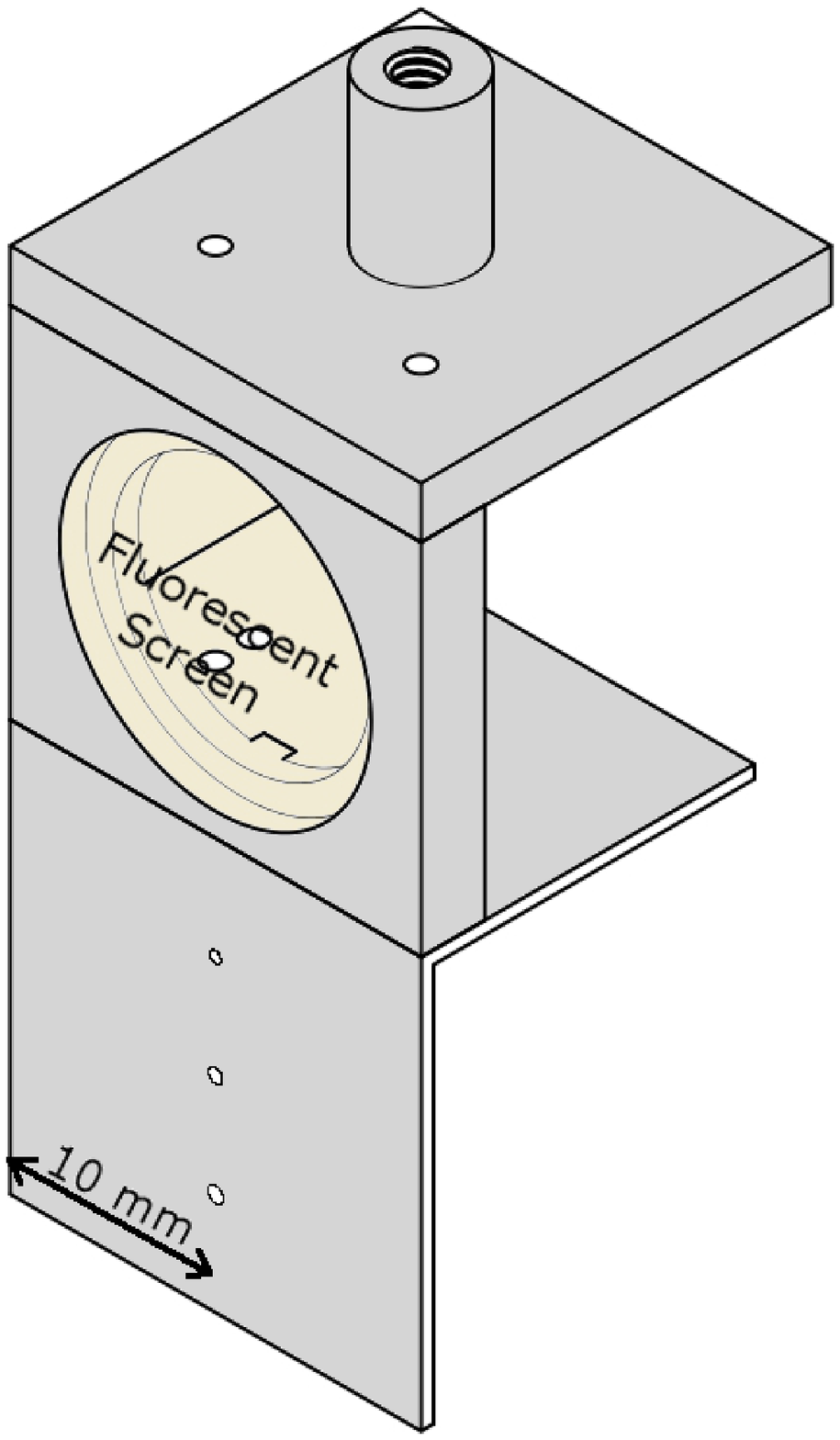}
\caption{Left: Schematics of the first few elements in the
straight section of UMER: electron gun, plane of aperture disk
(AP1), solenoid lens, diagnostic chamber 1 (IC1) with window,
plane of second aperture (AP2), actuators, and first two
quadrupole lenses. The top actuator houses fluorescent screen
diagnostics and AP2. Right:  New aperture plate (AP2) attached to
the bottom of the fluorescent-screen cube - 3 small apertures are
shown.}
\end{minipage}%
\end{figure}

\section{LOW SPACE-CHARGE INTENSITY BY DOUBLE COLLIMATION AND SOLENOID
FOCUSING}
Figure 1, left, shows the schematics of the first few
elements downstream of the electron gun in UMER. We employ two
sets of apertures, AP1 and AP2, and solenoid focusing.  As shown
in Figure 1, right, the cube that houses fluorescent screen
diagnostics in IC1 is fitted at the bottom with a plate containing
3 apertures of varying radii. After viewing the beam and then
adjusting its size with solenoid focusing, it is possible to
accurately position one of the apertures to obtain small beam
currents. The plot in Figure 2, left, illustrates the calculated
envelope of the 6 mA (nominal) beam over 30 cm from AP1 to AP2 for
seven values of solenoid current. It is possible to over-focus the
beam to yield even larger beams onto AP2 than with zero solenoid
current. Figure 2, right, displays the beam envelope from AP1 to
AP2 when the solenoid current is 12 A.

The 6 mA beam expands freely (solenoid current = 0 A) from a
radius of 0.875 mm at AP1 to 6.7 mm at AP2. Thus, a second
aperture AP2 of radius 0.67 mm yields a new current of 50 $\mu$A
past AP2, with an estimated rms normalized emittance equal to
0.073 $\mu$m. The emittance of the apertured beam is obtained from
simulations with the WARP code \cite{Friedman14}; it can be
estimated also by simply multiplying the experimental 6 mA
emittance (0.75 $\mu$m) by (0.67/6.7). The experimental emittance
of the 50 $\mu$A beam, however, is smaller, 0.05 $\mu$m. It is
obtained via a standard quadrupole scan whereby the 2$\times$rms
beam radii squared (at a screen 40 cm downstream from AP2) are
plotted as a function of quadrupole (iQ1) current; the emittance
is extracted from the parabolic fit constants.

Changing solenoid focusing to vary the beam radius at AP2 yields
varying SC intensities for beam transport in UMER (SC intensity as
defined for a matched beam in a periodic lattice - see equation (2),
and the more general formula in \cite{Bernal16}.) Figure 3, left,
shows results of beam radius at AP2 from experiment, solutions of
the K-V envelope equations \cite{Reiser08,Bernal16}, and WARP
simulations, as a function of solenoid current. The WARP
simulations employ an initial semi-Gaussian distribution at AP1,
40,000 macro-particles and a solenoid field that includes terms to
fifth order. There is very good agreement between calculations and
experiment for solenoid currents from 0 to 7 A, and between WARP
simulations and envelope calculations for all solenoid currents.

\begin{figure}
\begin{minipage}[c]{0.42\linewidth}
\includegraphics[width=\linewidth]{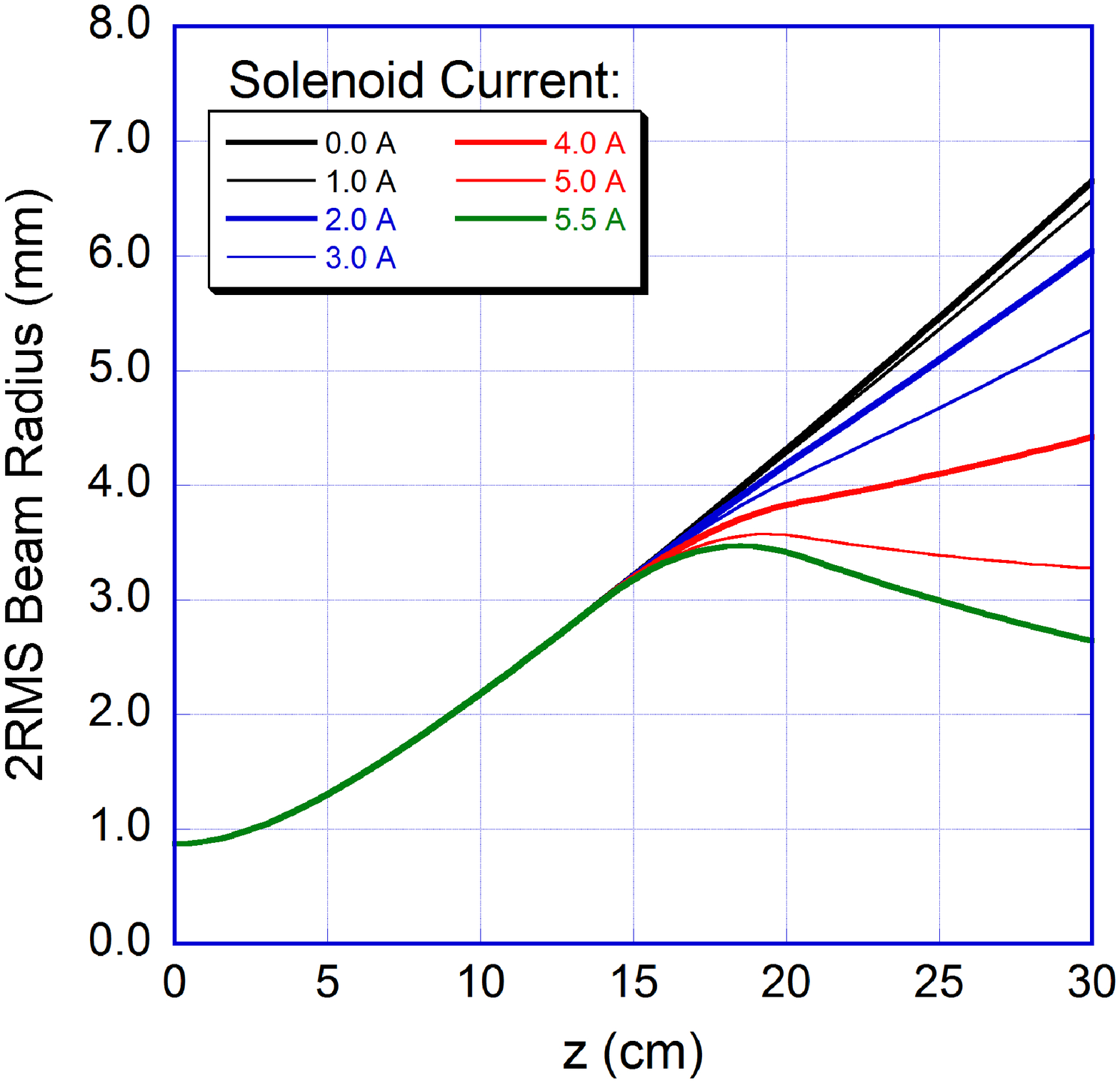}
\caption{}
\end{minipage}
\hfill \hspace{0.3in}
\begin{minipage}[c]{0.42\linewidth}
\includegraphics[width=\linewidth]{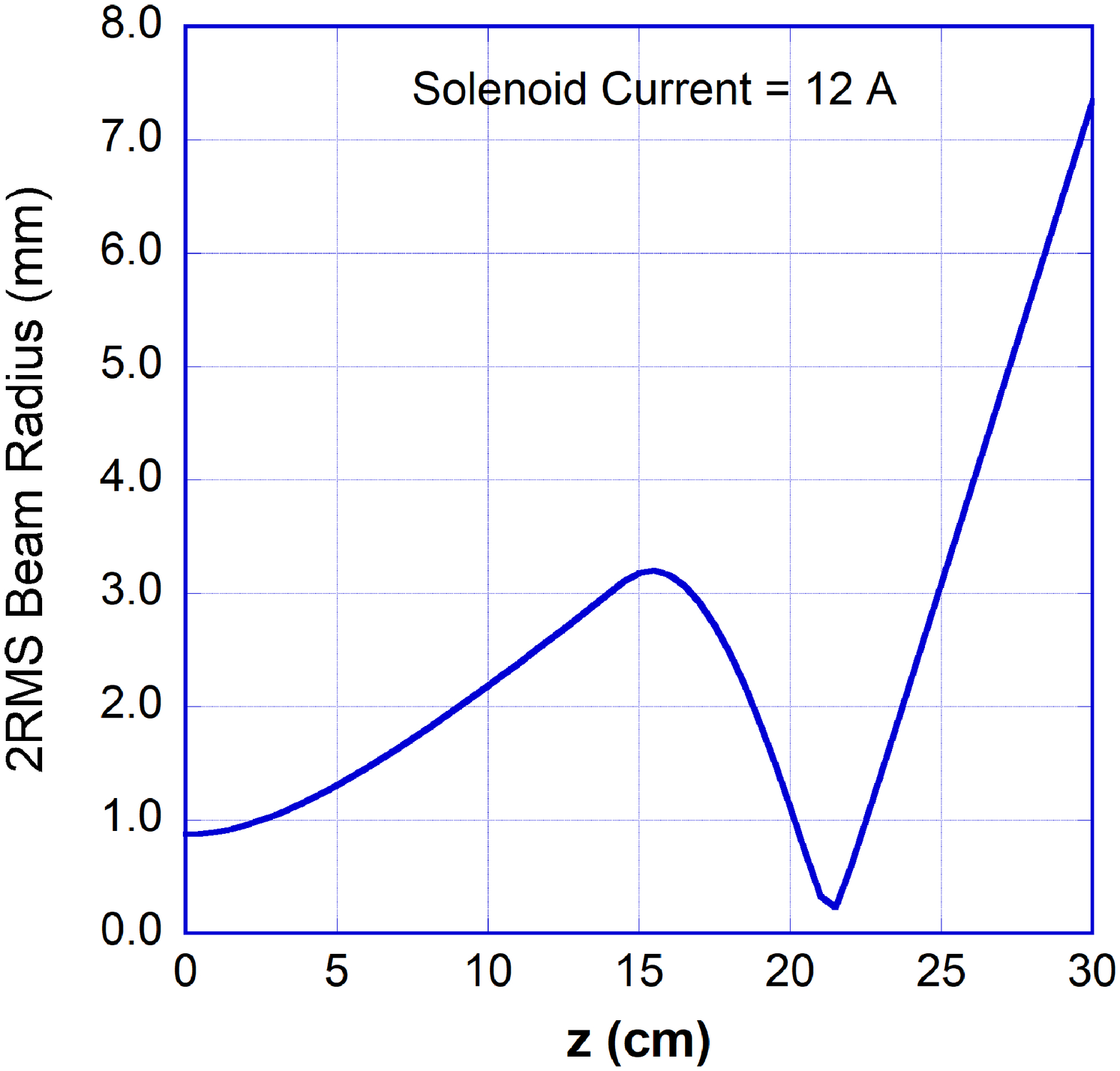}
\caption{Left: evolution of 2$\times$rms beam envelope radius from
AP1 to AP2 (see Fig. 1) for 7 values of solenoid current for the
6.0 mA (nominal) beam. Right: evolution of the envelope with
solenoid over-focusing. The short solenoid is located at $z =$
17.6 cm from AP1.}
\end{minipage}%
\end{figure}

\begin{figure}
\begin{minipage}[c]{0.42\linewidth}
\includegraphics[width=\linewidth]{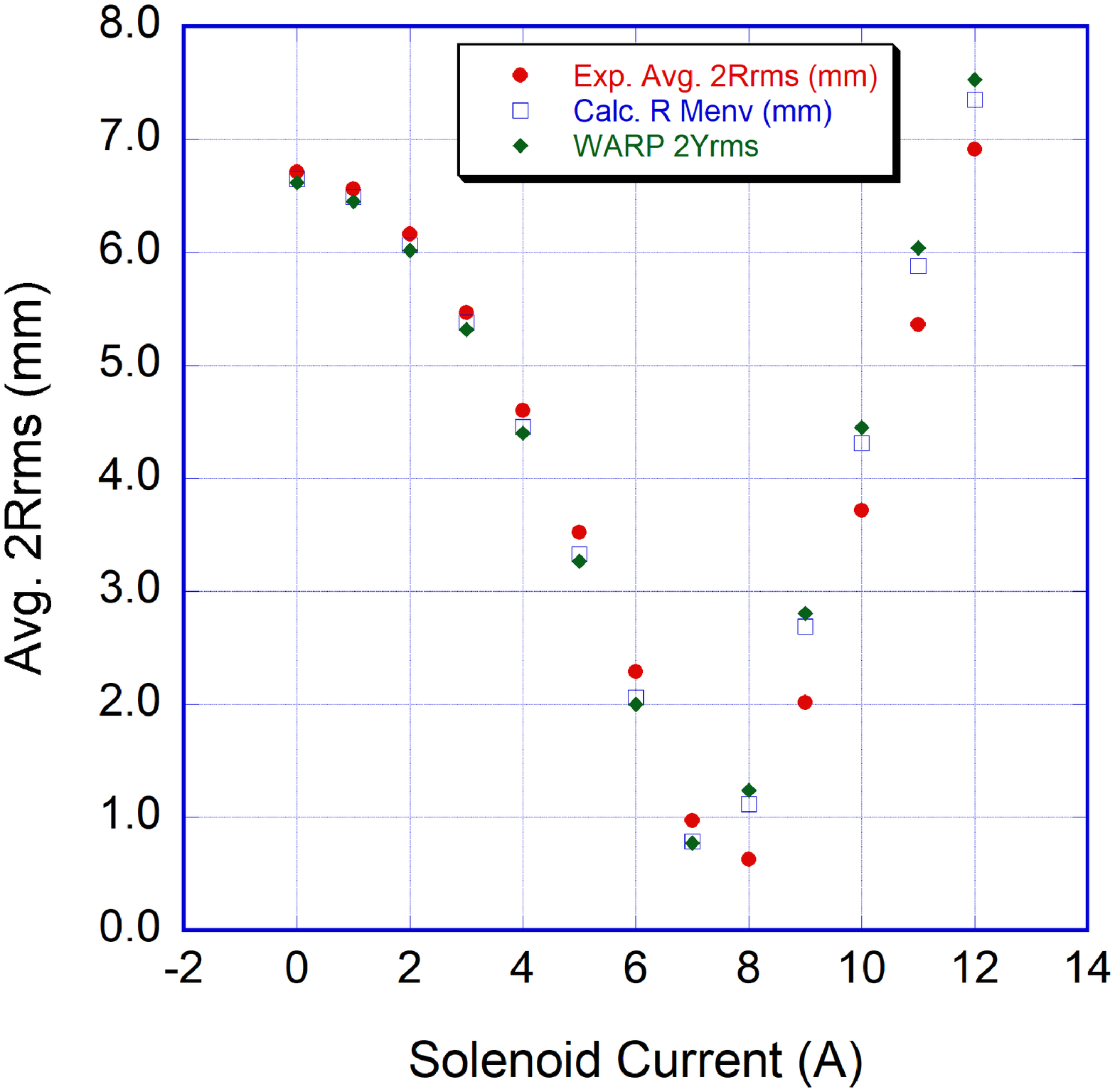}
\caption{}
\end{minipage}
\hfill \hspace{0.2in}
\begin{minipage}[c]{0.47\linewidth}
\includegraphics[width=\linewidth]{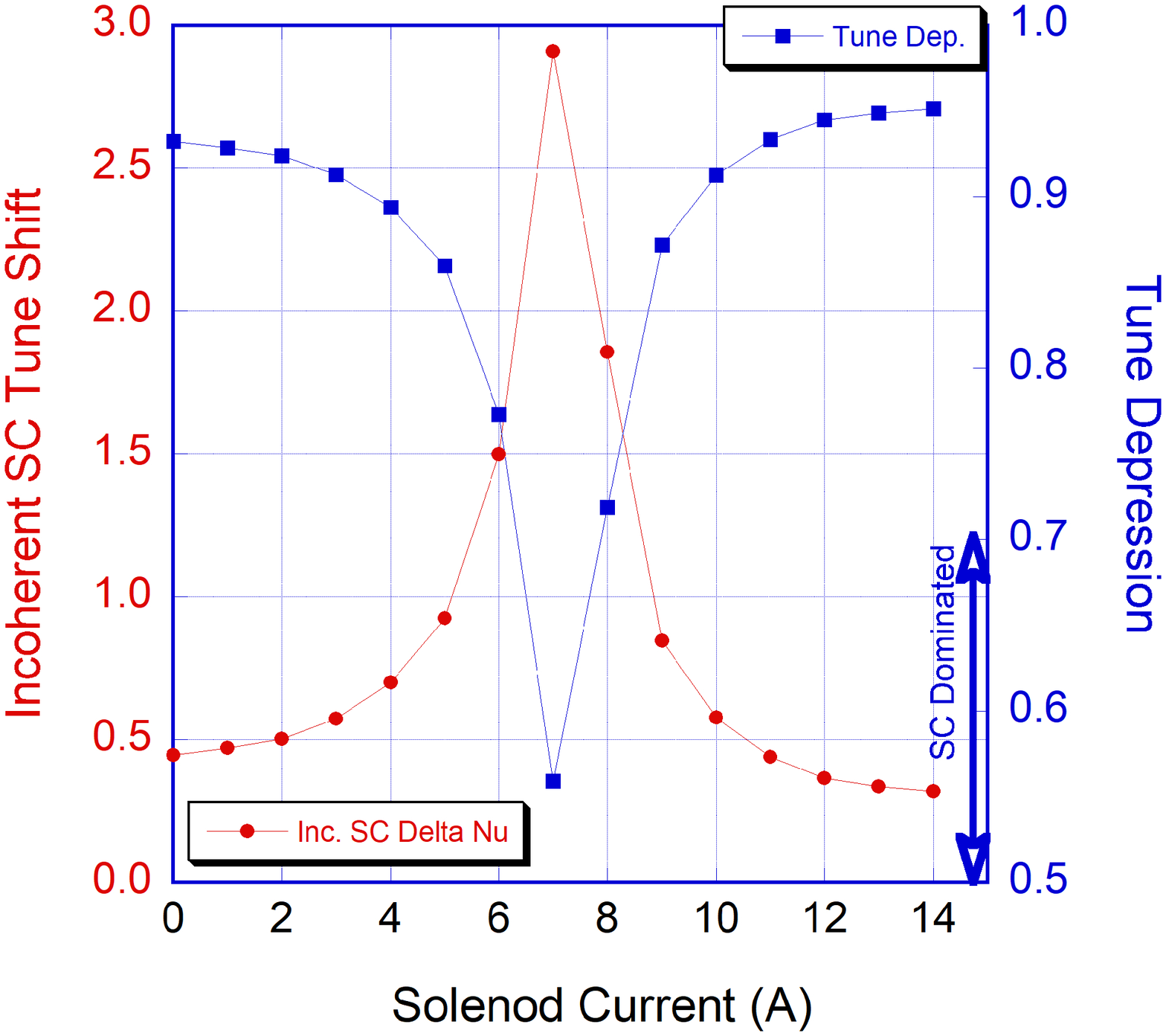}
\caption{Left: Beam radius at upstream plane of aperture AP2 as a
function of solenoid current (see Fig. 1); results from
experiment, envelope calculations and simulations with the WARP
code are shown. Right: Direct space charge parameters, tune shift
(left axis) and tune depression (right axis). The arrow on the
tune depression scale indicates the region of SC dominated
transport (only one point is clearly in that region - see also
table 3.)}
\end{minipage}%
\end{figure}

\begin{figure}
\begin{minipage}[c]{0.3\linewidth}
\includegraphics[width=\linewidth]{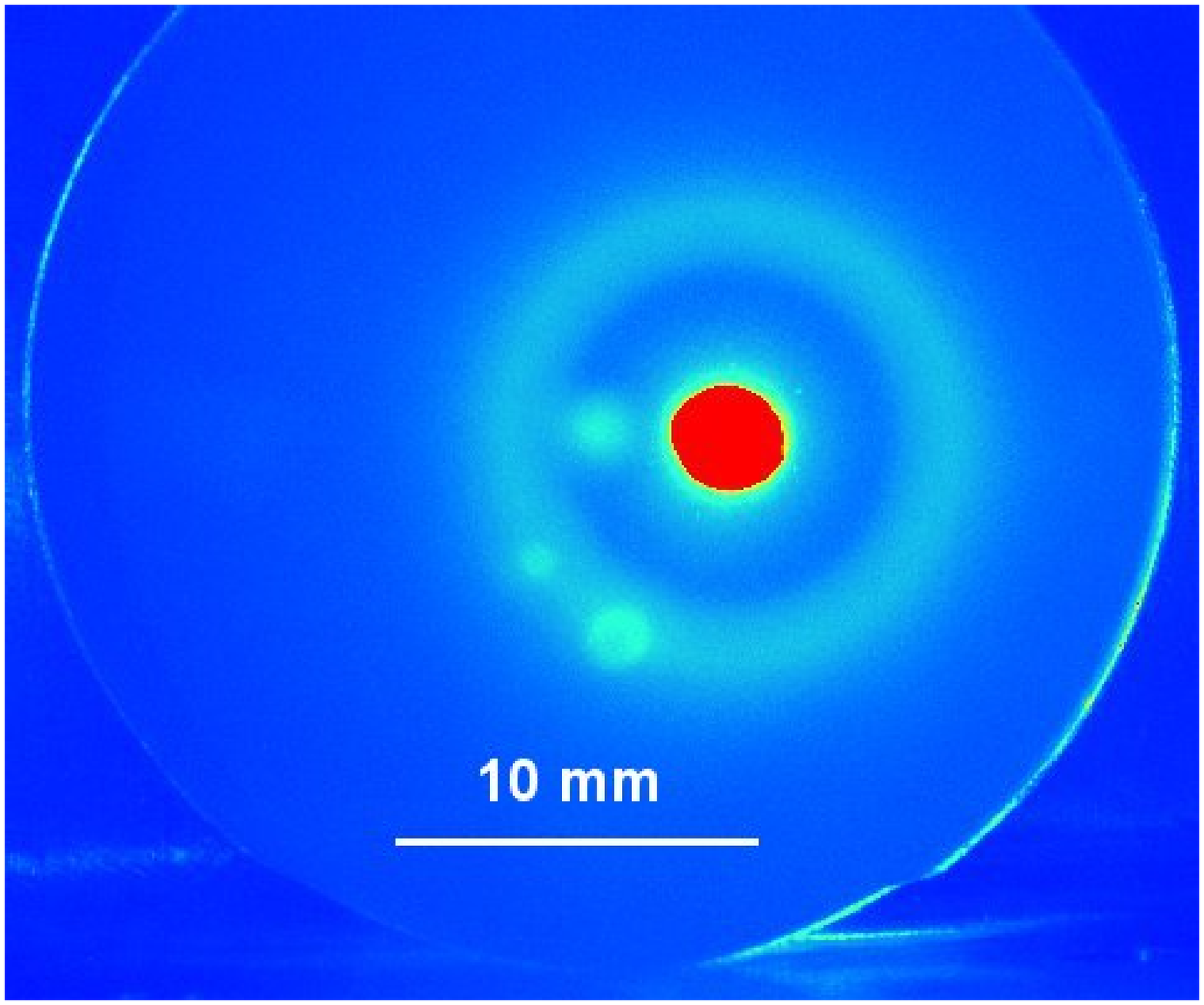}
\caption{}
\end{minipage}
\hfill \hspace{0.15in}
\begin{minipage}[c]{0.3\linewidth}
\includegraphics[width=\linewidth]{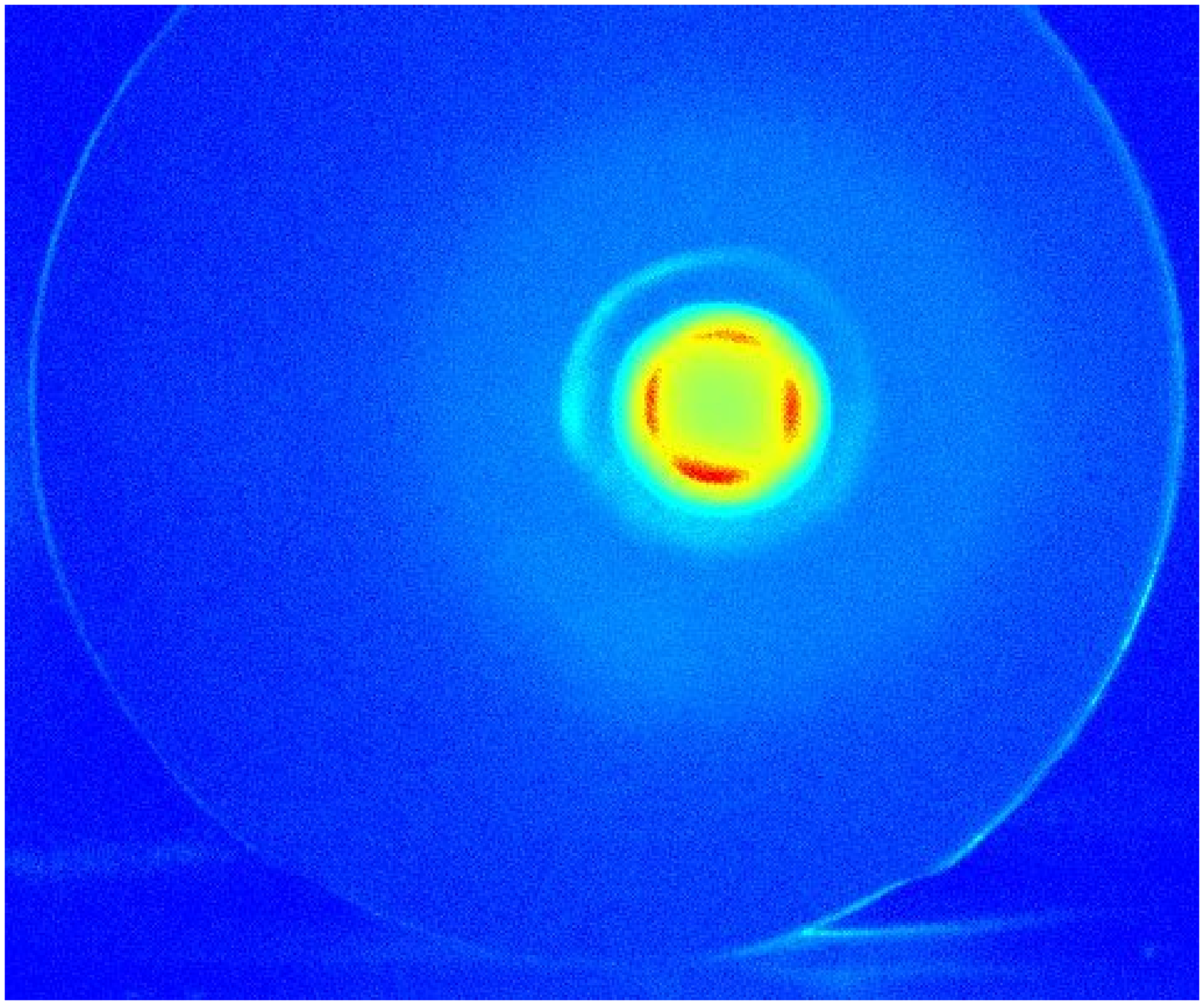}
\caption{}
\end{minipage}
\hfill \hspace{0.15in}
\begin{minipage}[c]{0.3\linewidth}
\includegraphics[width=\linewidth]{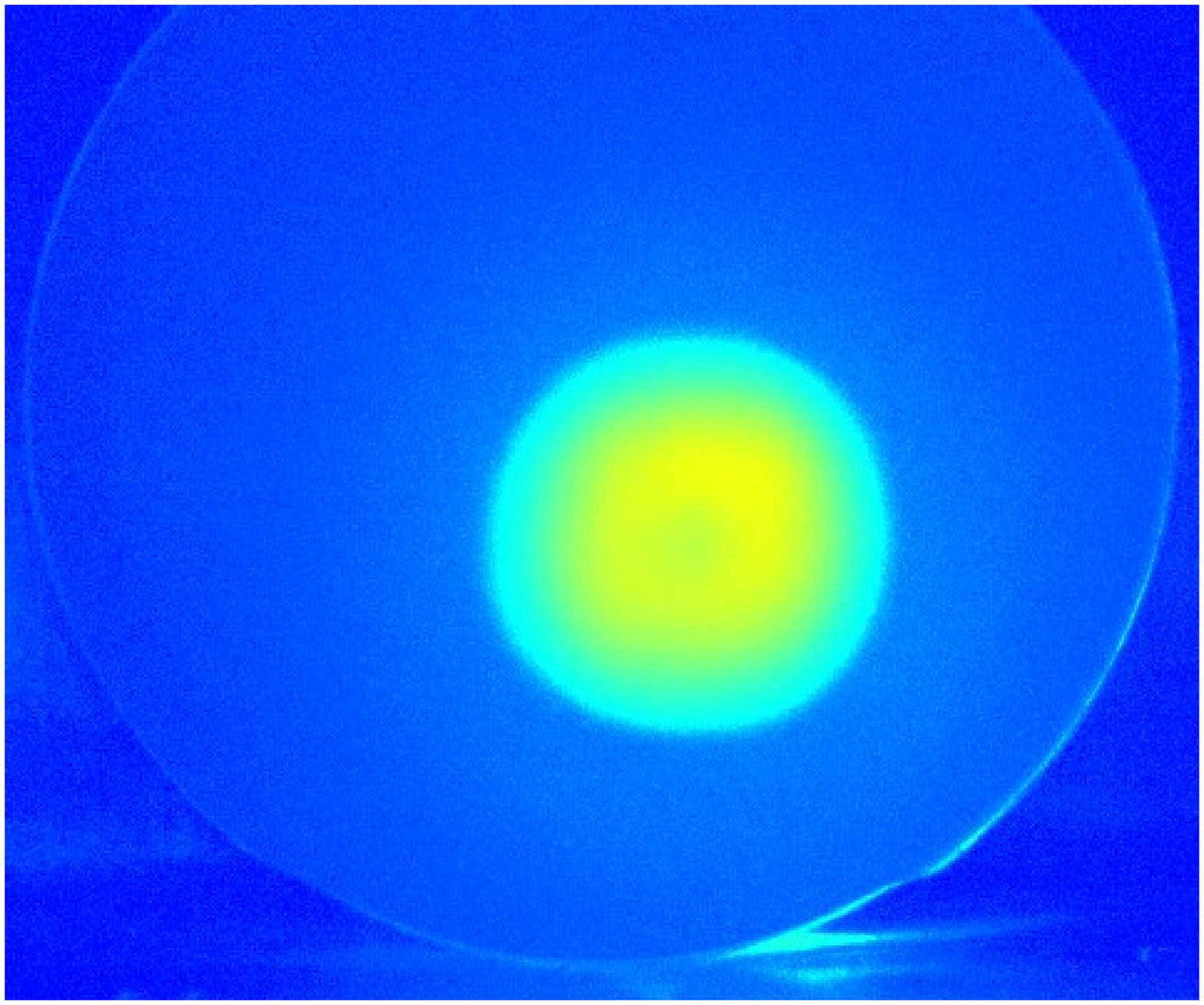}
\caption{Pseudo-color renderings of fluorescent screen beam
pictures at upstream plane of AP2 (see Fig. 1, right) for a
solenoid current of 7 A (left), 6 A (middle), and 10 A (right).
Both the halo and ghost images for the 7 A case (left) are from
the camera optics and/or fluorescent screen; the halo for the 6 A
case is deemed real.}
\end{minipage}%
\end{figure}

The variation in SC intensity is illustrated in Figure 3, right,
where both the incoherent SC tune shift (left axis) and the tune
depression (right axis) are plotted as a function of solenoid
current. The operating tune in UMER for these calculations is
$\nu_0$ = 6.6. The effective emittances past AP2 are extracted
from WARP simulations, while the beam currents are the initial
beam current scaled by the ratio (0.67 mm/$\textnormal{beam radius
at upstream plane of AP2 (mm)})^2$. Only one point in Fig. 3,
right, at solenoid current of 7 A, corresponds to SC-dominated
beam transport for which the tune depression is $<$ 0.71; the
point at 8 A is borderline between SC- and emittance-dominated
transport. At a solenoid current of 7 A, the effective beam radius
at AP2 is close to 1 mm (Fig. 3, left), leading to a beam current
of near 3 mA past AP2 and an emittance smaller but close to the
original one for the 6 mA beam. Note also that over-focusing the
beam (solenoid current $>$ 10 A) leads to slightly smaller SC tune
shifts, and tune depressions closer to 1.0. Table 3 summarizes the
cases discussed so far and includes the dimensionless SC intensity
parameter $\chi$, which is $<$ 0.5 for emittance-dominated
transport. The table also contains partial experimental data of
beam current and emittance; we are working on solving alignment
issues with the solenoid to extend the measurements.

Figure 4 shows pseudo-color versions of fluorescent screen beam
pictures at AP2 for three cases of solenoid current: 7 A, 6 A, and
10 A (see also Fig. 3). The WARP simulations do not reproduce the
halos seen at 6 and 7 A, indicating that their source, if real, is
not from solenoid spherical aberration, which is included in the
solenoid model. However, the relative intensities and shapes of
the halos lead us to believe that only the halo for the 6 A case
is real: the large halo is as dim as the round ghost images (about
1 -- 2 out of 255 for 8-bit grayscale pictures) and perfectly
formed. The actual halo would be blocked by the bottom aperture at
AP2 (radius = 0.67 mm), but it could reappear downstream leading
to an rms emittance significantly larger than calculated at a
solenoid current of 6 A.

In conclusion, double collimation and solenoid focusing can be
used to vary the SC intensity and reduce the incoherent tune shift
from 2 - 3 to $<$ 0.5, and change the tune depression from 0.55 to
$>$ 0.9. Naturally, using the middle and top apertures at AP2
(Fig. 1, right), which are smaller than the one used for
calculations and measurements for this work, could lead to
additional reduction of direct SC, but S/N issues arise because of
the small currents thus obtained.

\begin{table}[h]
\caption{Effect of solenoid focusing on beam and space-charge
parameters with 6 mA at AP1, and AP2 radius = 0.67 mm. The
operating tune in UMER is $\nu_0$ = 6.6.}
\tabcolsep7pt\begin{tabular}{ccccc} \hline
\tch{1}{c}{b}{Solenoid\\ Current} & \tch{1}{c}{b}{Beam Current, RMS\\ Emittance (morm.)} & \tch{1}{c}{b}{$\nu/\nu_0$,\\ $\Delta\nu$} & \tch{1}{c}{b}{SC Intensity \\$\chi$} \\
\hline
0 A  & 49.6 (58)$^*$ $\mu$A, 0.073 (0.05)$^*$ $\mu$m & 0.93, 0.45 & 0.130 \\
2  & 60.0 (63)$^*$ $\mu$A, 0.078  & 0.92, 0.50 & 0.146 \\
4  & 112 (99)$^*$ $\mu$A,  0.103  & 0.89, 0.70 & 0.201 \\
7  & 3.70 mA,     0.615  & 0.56, 2.91 & 0.687 \\
8  & 1.41 mA,     0.434  & 0.72, 1.86 & 0.483 \\
12 & 68.6 $\mu$A, 0.070  & 0.94, 0.36 & 0.107 \\
\hline
\end{tabular}
\end{table}
\vspace{-.25in}
\begin{quote} \begin{quote} \begin{quote}
{\small
  $^*$Measured (in parenthesis). }
\end{quote} \end{quote} \end{quote}

\newpage
\section{ACKNOWLEDGMENTS}
This work was supported by the Office of Science, Office of High
Energy Physics of the U.S. Department of Energy.



\end{document}